\documentclass[]{aa}
\usepackage{times}
\usepackage{graphics}
\usepackage{epsf}
\begin{document}

\title{Dust emission from inhomogeneous interstellar clouds: \\
radiative transfer in 3D with transiently heated particles
}

\titlerunning{Dust emission from inhomogeneous interstellar clouds}

\author{M.~Juvela \inst{1}
\and P.~Padoan \inst{2}}
\offprints{M.~Juvela}
\institute{
$^1$Helsinki University Observatory, T\"ahtitorninm\"aki, P.O.Box 14,
SF-00014 University of Helsinki, Finland (mjuvela@astro.helsinki.fi) \\
$^2$Jet Propulsion Laboratory, 4800 Oak Grove Drive, MS 169-506, California
Institute of Technology
}  

\date{Received <date> ; accepted <date>}

\maketitle

\abstract{
Due to the complexity of their structure, the theoretical study of
interstellar clouds must be based on three-dimensional models.
It is already possible to estimate the distribution
of equilibrium dust temperature in fairly large 3D models and, therefore, also
to predict the resulting far-infrared and sub-mm emission. Transiently heated
particles introduce, however, a significant complication and direct
calculation of emission at wavelengths below 100$\mu$m is currently not
possible in 3D models consisting of millions of cells.
Nevertheless, the radiative transfer problem can be solved with some
approximations. We present a numerical code for continuum radiative transfer that is
based on the idea of a `library' describing the relation between the intensity
of the local radiation field and the resulting dust emission spectrum. Given
this mapping it is sufficient to simulate the radiation field at only a couple
of reference wavelengths. Based on the library and local intensities at the
reference wavelengths, the radiative transfer equation can be integrated
through the source and an approximation of the emission spectrum is obtained.
Tests with small models for which the radiative transfer problem can be solved
directly show that with our method, one can easily obtain an accuracy of a few
per cent. This depends, however, on the opacity of the source and the type of
the radiation sources included.
As examples we show spectra computed from three-dimensional MHD simulations
containing up to 128$^3$ cells. The models represent starless, inhomogeneous
interstellar clouds embedded in the normal interstellar radiation field.
The intensity ratios between IRAS bands show large variations that follow the
filamentary structure of the density distribution. The power law index of the
spatial power spectrum of the column density map is -2.8.
In infrared maps temperature variations increase the power at high spatial
frequencies, and in a model with average visual extinction $\langle A_{\rm V}
\rangle\sim$10 the power law index varies between -2.5 and -2.7.
Assuming constant dust properties throughout the cloud, the IRAS ratio $\langle
I_{\rm 60}/I_{\rm 100}\rangle$ decreases in densest cores only by a factor of $\sim$4
compared with the value in diffuse medium. Observations have shown that in
reality the ratio can decrease twice as much even in optically thinner clouds.
This requires that most of the small grains are removed in these regions, and
possibly a modification of the properties of large grains.

\keywords{ISM: clouds -- Infrared: ISM -- Radiative transfer }
}

\section{Introduction}

Present knowledge on the large scale infrared emission from interstellar dust
is based largely on the all-sky surveys performed by the IRAS and the COBE
satellites. IRAS was the first to observe the sky at a resolution of a few arc
minutes at wavelengths between 12 and 100$\mu$m. Clear variations were
observed in the infrared spectrum of interstellar clouds, e.g. the ratio of
60$\mu$m vs. 100$\mu$m dropping towards dense clouds (e.g. Laureijs et al.
\cite{laureijs96}; Abergel et al. \cite{abergel94}). These were interpreted
as the result of abundance variations between classical, large grains and very
small dust particles that are transiently heated above the equilibrium
temperature (e.g. D\'esert et al. \cite{desert90}). Emission in the
mid-infrared and especially the IRAS 12$\mu$m band, are now believed to be
caused by even smaller grains or very large molecules, the most common
candidates being Polycyclic Aromatic Hydrocarbons (PAHs) (L\'eger et al.
\cite{leger89}; D\'esert et al. \cite{desert90}; Li \& Draine \cite{li01}).
The PAH emission traces the warmer outer surfaces of the clouds.

The COBE/DIRBE instrument covered a wider wavelength range from near-infrared
up to 240\,$\mu$m and the FIRAS instrument up to 1\,cm. The spatial resolution
was, however, comparatively poor i.e. $\sim$40$\arcmin$ for DIRBE and
$\sim$7$\degr$ for FIRAS. Nevertheless, the extended wavelength coverage
revealed surprisingly large quantities of cold dust that were not observed
with IRAS. Detailed analysis of the COBE data has shown that at least two
emission components are needed to explain the observed far-infrared spectra.
For most of the diffuse regions the dust emission follows a modified black
body law with temperature $T\sim$17.5\,K (Boulanger et al.
\cite{boulanger96}; Dwek et al. \cite{dwek97}). Lagache et al. (\cite{lagache98})
found, however, a colder component ($T\sim15$\,K) associated with molecular
clouds. Finkbeiner et al. (\cite{finkbeiner99}) reached similar conclusions
but included an even colder, $T\sim$9\,K, component.

Recent observations with the PRONAOS balloon borne observatory have confirmed
the lowering of the colour temperature towards molecular clouds and, e.g. in
the case of the Polaris flare ($A_{\rm V}\sim$1), towards moderately dense regions
(Stepnik et al. \cite{stepnik_esa}). 
Similar effects have been seen in ISOCAM
studies towards cirrus like clouds ($A_{\rm V}\sim$0.5; Miville-Desch\^enes et al.
\cite{miville02}) where the temperature variations caused by extinction are
quite insignificant. The observed increase in the sub-mm emission can be
explained by grain growth and the formation of large dust aggregates (Stepnik
et al. \cite{stepnik_esa}; Cambr\'esy et al. \cite{cambresy01})
and in dense regions the very small grains seem to have disappeared almost
completely. Conversely, in the more diffuse medium, large quantities of small
grains can be produced by grain shattering (see e.g. Miville-Desch\^enes et
al.
\cite{miville02}).

Based on these observations and theoretical studies we have obtained a rough
picture of the dust properties and infrared emission from interstellar clouds.
In the outer layers we have the PAH emission and the warm, very small dust
grains. Deeper in the clouds the extinction of short wavelength radiation
reduces mid-infrared emission. At the same time, the grains start to form
larger aggregates and/or acquire mantles as gas molecules freeze onto them.
This increases the far-infrared and sub-mm emissivity relative to the
absorption at short wavelengths and the physical grain temperatures also drop.
The colour temperature therefore drops to approximately 10\,K (or possibly
even lower) in dense cloud cores (e.g. Juvela et al.~\cite{juvela02a}).

Theoretical and laboratory studies have provided us with some information on the
properties of likely dust grain materials. Together with radiative transfer
calculations, these have made it possible to determine size
distributions for the different dust components consistent with the emission
properties (e.g. D\'esert et al. \cite{desert90}; Li \& Draine
\cite{li01}). On the other hand, radiative transfer models can be used to
study the variations of these properties in individual objects. These previous
examples show that the dust emission spectrum is variable over the whole range
from near-infrared to sub-mm. The changes are related to the local radiation
field and the local density. So far, modelling has concentrated either on the
determination of the dust emission under given radiation conditions, or at
most in one-dimensional, spherical models divided into some tens of cells.

In the related field of the modelling of molecular line emission,
three-dimensional models have already been used for some time (Park \& Hong
\cite{park95}; Park et al. \cite{park96}; Juvela \cite{juvela98}; Padoan et al.
\cite{padoan98}; Juvela et al. \cite{juvela01}). In that field,
inhomogeneous source structure has an even stronger effect on the emerging
radiation and, on the other hand, direct calculations are already possible for
models with up to $\sim$100$^3$ cells or, with approximate methods, even
higher (Ossenkopf \cite{ossenkopf02}).

The continuum emission is not as directly linked with the local, physical
conditions. It is, nevertheless, affected by the variations in the radiation
field (extinction of the external radiation field and internal sources) and
the changes in the dust properties. So far dust emission has not been studied
with general three-dimensional cloud models. The main reason is the complexity
caused by the transiently heated small dust particles. In order to be able to
predict the emission accurately, the distribution of the grain temperatures
must be first determined in each computational cell. Each dust population must
be discretized into separate grain size intervals. For each of these, a couple
of hundred enthalpy bins are needed and the solution of the associated set of
linear equations takes typically several seconds. This becomes significant
when the model contains millions of cells and, in practice, it limits
direct calculations to one- or possibly two-dimensional models.

We will discuss how this limitation can be overcome by suitable approximations
without siginificantly affecting the accuracy of the results. This will be
based on the assumption that the local spectral energy distribution can be
deduced from the intensity at suitably selected reference wavelengths. One
must first determine a mapping between these intensities and the resulting
local spectral energy distribution. The radiative transfer problem needs only
to be solved at the reference wavelengths, and by using the established mapping,
one can solve the radiative transfer problem. The solution is approximative
but can be be made sufficiently accurate within the limitations of present day
computers. The method makes it possible to study explicitly the effects that
the cloud structure has on the infrared emission, and the method could be
useful e.g. in the simulation of the spectral energy distribution of galaxies.

We first explain the implementation of the radiative transfer program
(Sect.~\ref{sect:program}) and discuss its accuracy
(Sect.~\ref{sect:tests}). In Sect.~\ref{sect:models} we apply the method to
three-dimensional model clouds that are based on numerical simulations of supersonic
magneto-hydrodynamic (MHD) turbulence and we show
some observable consequences of the inhomogeneity of the clouds. Finally, in
Sect.~\ref{sect:adjust} we discuss some qualitative effects of spatial
variations in the dust properties.

\section{The program} \label{sect:program}

\subsection{Direct solution} \label{sect:basic}

The full radiative transfer problem can be divided into two largely
independent parts. In the first one determines the intensity of the radiation
field at each position of the cloud (in our case in each discrete cell of
finite volume). In the second part this information and the selected dust
model are used to derive the local dust temperature and emission.
In interstellar clouds dust emission mainly takes place in the infrared part
of the spectrum. As long as the external radiation field is relatively weak
the dust emission has only a small effect on the heating of grains (see e.g.
Hollenbach et al. \cite{hollenbach91}; Bernard et al. \cite{bernard92}). This
is true for all models considered in this paper and, therefore, the two parts
of the computation mentioned above can be carried out sequentially without
iteration. 
In the following we will ignore other processes affecting the dust
temperature. For example, the gas-dust coupling becomes important only at
densities above $\ga$10$^5$\,cm$^{-3}$ (e.g. Kr\"ugel \& Walmsley
\cite{krugel84}) and the clouds considered here have much lower densities. 

Our three-dimensional model clouds are divided into cubic cells, each with a
constant density. First, the optical depths for absorption and scattering as
well as the parameters of the scattering function are calculated for each
cell. The usual Henyey-Greenstein formula (\cite{henyey41}) and the asymmetry
parameter $g$ are used to define the scattering function. The radiation field
is calculated with Monte Carlo methods. Model photons are initiated at random
locations at the cloud boundary in order to simulate the background radiation,
and within the cloud to simulate emission from the dust and possible internal
sources. The path of the model photon is tracked in the cloud. A model photon
represents a number of real photons. Absorbed photons are counted in each cell
that is crossed, and from time to time, the model photon is scattered towards a
new direction as determined by the dust model. Because of the frequency
dependence of the scattering probability, the simulation has to be carried out
separately for each frequency.

The simulation determines the spectrum of the incoming radiation for each
cell. In the second phase this is used to compute the dust temperature. The
dust model consists of several dust populations, and each population is
further discretized according to grain size. For each of these we have a
probability distribution of the grain enthalpy, discretized usually to a few
hundred bins. The transition probabilities for cooling processes are
calculated beforehand, based on the specific heat and absorption cross
sections. The transitions upwards in energy are obtained by integration from
the simulated intensity spectrum. The formulae we use are exact only for a
probability distribution that is flat a function of grain energy
(temperature). We have also implemented the `thermal discrete' method
described by Draine \& Li (\cite{draine01}) and have tested that our method
gives, with the used discretizations, identical results for the studied
wavelength range. Once the number of grains as the function of temperature
(enthalpy) is known the local emission can be calculated.

The two steps still need to be iterated, if the dust emission itself
contributes to the grain heating. This requires, however, either the presence
of strong, local heating sources or very high extinction. Bernard et al.
(\cite{bernard92}) studied the case of spherical clouds heated by normal ISRF
(a slightly modified version of the ISRF given by Mathis et al.
(\cite{mathis83})).
The IR heating was found to be insignificant for visual
extinctions $A_{\rm V}<100^{\rm m}$ through the cloud, and even with
$A_{\rm V}=1000^{\rm m}$, it increased the sub-mm emission only by $\sim$5\%. The
effect is due to an increase in the temperature of the large, classical
grains. Shorter wavelength emission is in practice not affected, since the
infrared photons emitted by the dust grains are not energetic enough to cause
significant transient heating of the smaller grains.

\subsection{The library method} \label{sect:approx}

For large, three-dimensional models consisting of $>100^3$ cells the described
direct approach is no longer possible. Simulation of the necessary number of
frequencies ($>100$) could still be done, but it would possibly take days on a workstation,
depending on the required accuracy. The main problem is the solution of the linear
equations. Each dust population is discretized into ten
or more size intervals, each further divided into possibly several hundred
enthalpy intervals. The solution of the associated linear
equations for the transition probabilities requires more than a second per cell,
or a month for a model with $100^3$ computational cells. The direct approach is clearly
limited to very small models (or very large supercomputers).

The spectrum of the local radiation field is very similar in many parts of the
cloud. For example, in a homogeneous, semi-infinite plane or a homogeneous
sphere, the intensity and spectrum of incoming radiation depends only on one
parameter, the distance from the free surface. In these cases one could
calculate the dust emission at certain depths and obtain the emission from
intermediate positions with interpolation.

Inside an inhomogeneous or an irregularly shaped cloud there is more variation
in the spectral energy distribution of the incoming radiation. Usually, in a
homogeneous cloud, the extinction reduces radiation more strongly at short
wavelengths. This is not true, however, in the extreme case of inpenetrable
clumps. If clumps absorb all incoming radiation irrespective of its
wavelength, the intensity inside the cloud will be reduced proportionally to
the solid angle blocked by the clumps, while the shape of the spectrum will
remain unchanged. Clearly, in a general case, the changes in the radiation
field cannot be tracked with just one parameter (e.g. visual extinction). We
argue, however, that for a given model, the spectrum can be sufficiently
constrained with knowledge of the intensity at possibly only two or three
reference wavelengths, $\lambda^{\rm ref}_i$.

The first step in our approximation is to find the range of intensities and
spectral energy distributions of the incoming radiation. For this purpose, we
reduce the discretization of the model down to $\sim 32^3$ cells. This allows
fast solution of the radiative transfer part of the problem. The result is the
simulated intensity in each cell. We take the range of intensities found at
the first reference wavelength, $\lambda^{\rm ref}_{\rm 1}$, and discretize
that into logarithmic intensity intervals. The procedure is repeated
recursively at the other reference wavelengths. The result is a tree structure
that represents a discretization of all observed intensity combinations,
($I$($\lambda^{\rm ref}_{\rm 1}$), I($\lambda^{\rm ref}_{\rm 2}$), {\ldots}). In the
leaves of the tree we store the average, full spectrum that corresponds to a
unique combination of intensities at the reference wavelengths. If the
reference wavelengths are suitably selected, the tree and knowledge of
intensities at reference wavelengths can be used to obtain an estimate of the
full spectrum.

\begin{figure}
\resizebox{\hsize}{!}{\includegraphics{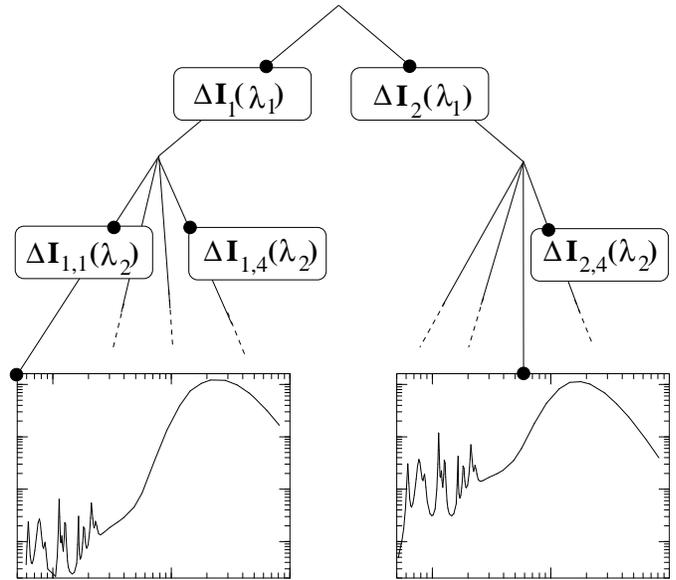}}
\caption[]{
A schematic view of the tree structure used in the library method. At each
level the intensity of the incoming radiation at one wavelength is examined
and each subnode corresponds to a different intensity interval, $\Delta I$. The
leaves of the tree correspond to a unique combination of intensity values at
the reference wavelengths and they hold the corresponding emission spectrum.
Once the intensity of the incoming radiation is known at reference
wavelengths, the emission can be read from the tree.}
\label{fig:tree}
\end{figure}

The second step of our approximation is the solution of the dust emission
corresponding to each spectrum of incoming radiation that was stored in the
leaves of the tree. This way, we obtain a mapping from the incoming intensity
at the reference wavelengths to the corresponding full spectrum of local dust
emission. We call this structure `the library' 
(see Fig.~\ref{fig:tree}).

To predict dust emission from a large, three-dimensional model, we first
simulate the radiation field at the small number of reference wavelengths,
$\lambda^{\rm ref}_i$. The radiative transfer equation can then be directly
integrated through the cloud, using in each cell the library to transform the
simulated values of incoming intensity at the reference wavelengths into the
full spectrum of emitted intensity. We call this `the library method'. When we
look up the emission spectrum from the library, we also perform linear
interpolation based on the values $I$($\lambda^{\rm ref}_{\rm i}$). Again, if absorbed
dust emission is important (e.g. close to strong sources or in very opaque
regions) iteration is needed.

\section{Tests of the library method} \label{sect:tests}

\subsection{The model clouds and the dust model}

We will now estimate the accuracy of the approximate method described in
Sect.~\ref{sect:approx}. The method is based on the discretization of the
incoming intensity into bins for which the corresponding dust emission
spectrum is only computed for the average spectrum. The accuracy can therefore
always be improved by selecting a finer discretization and/or including more
reference wavelengths. In order to obtain a significant advantage over the
direct solution, the total number of spectrum bins must, however, be much less
than the number of cells in the model.

In this paper, we limit the study to starless clouds heated by the
interstellar radiation field. We take as a test case an MHD simulation which
is the model $C$ from Juvela et al. (\cite{juvela01}). The model is both
supersonic and super-Alfv\'enic with acoustic and Alfv\'enic Mach numbers
$ {M}\approx{M}_{\rm A}\approx$10. The MHD simulation included self-gravity.

The dust model is taken from Li \& Draine (\cite{li01}) and includes
silicate grains (sizes $a$$>$3.5\AA), graphite grains ($a\ga$50\AA) and PAHs
(from $a$=3.5\AA to $\sim$50\AA). The PAH properties correspond to cold
neutral medium. The grain size distributions are the same as in Li \& Draine
(\cite{li01}). The only difference is the way PAH and graphite grain
populations are joined. Instead of a gradual change in the description of
absorption coefficient (Li \& Draine \cite{li01}, Eq. (2)) we only join the
grain size distributions smoothly around 40$\mu$m. 
The optical properties are taken from the files available through the website
of B. Draine. The original references are Laor \& Draine (\cite{laor93}) for graphite
and silicate grains and Li \& Draine (\cite{li01}) for the PAH properties.

At this point, we are interested in the differences between
the two methods (i.e. direct solution and the library method) and not in the
absolute accuracy. Therefore, the number of grain size intervals ($\sim$10),
grain energies ($<$300) and simulated frequencies ($\sim$100) were kept to a
minimum that still reproduces the Li \& Draine (\cite{li01}) spectra for
diffuse ISM within $\sim$10\% accuracy over the studied wavelength range,
5--300\,$\mu$m. The largest differences result from a different method used to
join the PAH and graphite grain populations around grain size 40$\mu$m. The
interstellar radiation field adopted is from Mathis et al. (\cite{mathis83}).

The original cloud model represented a cloud with a linear size $\sim$6\,pc
and a mean density of 320\,cm$^{-3}$. For these parameters, the average column
density is $N({\rm H}_2)$=5.9$\times$ 10$^{21}$\,cm$^{-2}$ and the
corresponding average visual extinction is $A_{\rm V}=$2.8~$^{\rm m}$. The cloud is
inhomogeneous and the extinction varies depending on the line of sight. In the
following, extinction and infrared spectra are only calculated toward one
direction. The range of extinctions is $A_{\rm V}=$0.31--24.6. Higher extinction
means larger variations in intensity. In our library method, this should lead
to larger errors, at least if the library always contains the same number of
intensity intervals.

We created two new models by scaling the
densities of the original model. For the radiative transfer problem, the
relevant parameter is the column density rather than the volume density and
the results of the tests are equally applicable to smaller and
correspondingly denser sources. The parameters of the models are listed in
Table~\ref{table:models}. The visual extinction is in all cases well below
1000$^{\rm m}$ and dust emission can be solved without iteration (see
Sect.~\ref{sect:basic}).

\begin{table}
\caption[]{Parameters of the model clouds used in the testing: name of the
model, average density assuming linear size $L$=6\,pc, and the average and
maximum values of visual extinction seen towards the direction for which the
dust emission was computed.}
\begin{tabular}{crrr}
model name  &  $\langle n\rangle$        &  $\langle A_{\rm V}\rangle$ &
$A_{\rm V}^{\rm MAX}$  \\
            &  [cm$^{-3}$]  &                &        \\
\hline
C0          &  320     &  2.76     &  24.6  \\
C1          & 1280     & 11.04     &  98.5  \\
C2          & 5120     & 44.16     & 394.0  \\
\end{tabular}
\label{table:models}
\end{table}

\subsection{The size of the library}

With discretization reduced to $32^3$ cells, the emission spectra can be
calculated directly within a reasonable time, and comparison with spectra
computed with the method introduced in Sect.~\ref{sect:approx} makes it
possible to directly estimate the errors caused by the library method.  The
radiation field is simulated using the same set of random numbers, and
therefore the quoted errors reflect only the accuracy of the approximation.
The incoming intensity is estimated with Monte Carlo methods, and the
resulting random noise causes additional uncertainties. However, although the
simulation of the radiation field represents a rather small part of the total
computation time in our runs, the expected noise at individual frequencies is
only a few per cent. The effect on the dust emission is expected to be
somewhat smaller, since the computation of dust temperatures involves averages
over a number of frequencies. We computed maps of 32$\times$32 spectra using
the 32$^3$ cell models and either direct solution or the library method. We
first consider models without the PAH component, including only larger
carboneous grains and silicate grains.

With relatively low extinction, the model $C0$ is the easiest case.
With two reference wavelengths, 0.12$\mu$m and 9.1$\mu$m, both with 20
intensity intervals, the average relative error in the computed emission was
below 2\% in the range 5--300\,$\mu$m. Taking into account all spectra and
all wavelengths in this range, the maximum deviation between intensities
computed directly and with the library method was $\sim$6\%. This is very
encouraging, considering that in the model the emission per unit volume
($\sim$density) varies over nearly three orders of magnitude and the intensity
along different lines of sight over almost two orders of magnitude. The
variations in the emission per dust grain are also significant, but
much smaller, and this makes the use of the library method possible.

The main computational cost lies in the solution of the dust temperature
distributions, using the 20$\times$20 spectra determined by the discretization
of the intensity at the two reference wavelengths. Even for such a small
model, the computational time is significantly reduced, by a factor of $\sim
30^3/20^2\approx$68.

\begin{figure}
\resizebox{\hsize}{!}{\includegraphics{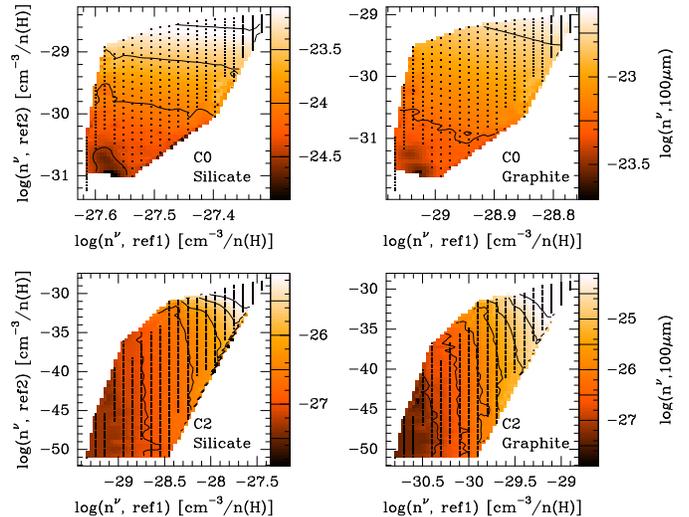}}
\caption[]{The 100\,$\mu$m emission as the function of the
local radiation field at reference wavelengths, $\lambda_{\rm ref1}=9.1\mu$m
and $\lambda_{\rm ref2}=0.12\mu$m. Intensities are given as the number of
photons absorbed per cubic centimeter, divided by hydrogen density. The upper
frames correspond to model $C0$ and the lower frames to model $C2$.}
\label{fig:parameters}
\end{figure}

Fig.~\ref{fig:parameters} shows the 100$\mu$m emission as the function of the
local intensity at two reference frequencies for model $C0$. In the plot the
contours are almost horizontal i.e. the emission can be predicted based on the
intensity at only the shorter of the two reference frequencies.
In models $C1$ and $C2$ the higher extinction makes the second reference
wavelength more important. Fig.~\ref{fig:parameters} shows clearly that in
model $C2$ the spectrum of the local radiation field can no longer be
approximated with only one parameter (e.g. visual extinction), and the
isocontours in the figure are not parallel to either axis. At low intensities,
i.e. deep inside the cloud, the emission follows the intensity at the longer
wavelength and the contours are closer to vertical. At higher intensities
(close to the cloud surface) the short wavelength radiation becomes relatively
stronger and it is more important in determining the dust temperature and the
infrared emission. As a result, the isocontours turn almost horizontal in
Fig.~\ref{fig:parameters}.

With the previous number of intensity intervals in the library, 20$^2$, the
rms errors increase to 3.5\% for model $C1$ and 6.2\% for model $C2$. The
errors might be reduced further by optimizing the selection of the reference
wavelengths. The wavelengths should obviously be well separated from each
other and correspond to sufficiently high optical depths so that intensity
variations are well above the Monte Carlo noise. The optimal values depend on
the radiation sources, the dust model and the cloud parameters. The reference
wavelengths used above are based on some experimentation but are probably not
optimal.

In optically thick clouds, the inclusion of a third reference wavelength may be
necessary. It is also better to use longer reference wavelengths in order to
avoid the extremely optically thick regime where accurate sampling of the
radiation field becomes difficult. Using 20 intensity intervals at wavelengths
0.20, 1.7 and 30$\mu$m, the rms errors are reduced to 1.8\% and 2.3\% for
models $C0$ and $C1$, respectively. For the most opaque model, $C2$, the rms
error is 3.8\%. It is, however, still possible to increase the size of the
library by at least one order of magnitude, and the errors caused by the use
of the library method can be kept below the noise due to the Monte Carlo
sampling.

The inclusion of PAHs does not cause significant problems for the library
method. For the model $C1$, with a library with three reference wavelengths
and 23$^3$ intensity intervals, the rms error in intensities between 5\,$\mu$m
and 300\,$\mu$m becomes 2.3\%.

It is not possible to give general estimates of the accuracy of the
library method in more complicated situations. For example, the radiation
field could probably not be estimated with sufficient accuracy using only two
reference wavelengths if the model included two radiation sources with
different emission spectra. The local field would clearly depend on the
distance to both sources. Furthermore, in an inhomogeneous cloud the way each
spectrum is attenuated depends on the density field which modifies both the
intensity and the shape of the radiation spectrum. In principle, the accuracy
can always be improved by increasing the number of intensity intervals and/or
the number of reference wavelengths. The library method remains viable as long
as the number of entries in the library is significantly smaller than the number
of cells in the cloud. For example, with four reference wavelengths, each with
30 intensity intervals the number library entries is 8.1$\times$10$^5$ and
method could still be used for all models with more that $\sim100^3$ cells.

\subsection{Effect of cloud resolution on the created library}

When the library is first created, the model cloud is used only to estimate
the range of intensities present in the cloud. In the library, the
combinations of intensities at the reference wavelengths do not correspond
directly to any specific cells in the cloud. Therefore, the previous error
estimates should also apply when the same library is used to calculate spectra
from models with higher discretization. The only exception may be the coldest
spot of the cloud. Low resolution leads to smoothing of density peaks and as a
result, a higher resolution model may contain a few cells where the intensity
is lower than anywhere in the low resolution model. If the library does not
contain entries corresponding to these lowest intensities the spectrum
predicted for the coldest core will have a spectrum that is too warm (i.e.
the coldest spectrum of the low resolution model).

There are several reasons why the effect is not expected to be very
significant. The library represents a mapping between the intensity of the
local radiation field and the dust emission per unit density, independent of
the density. Comparing models with different discretization, the largest
effect on the emitted intensity comes directly from the density.  Furthermore,
since the radiative transfer equation is integrated over the whole line of
sight the relative statistical errors in the emerging intensity are smaller
than the relative errors in local emissivity. 

\begin{figure}
\resizebox{\hsize}{!}{\includegraphics{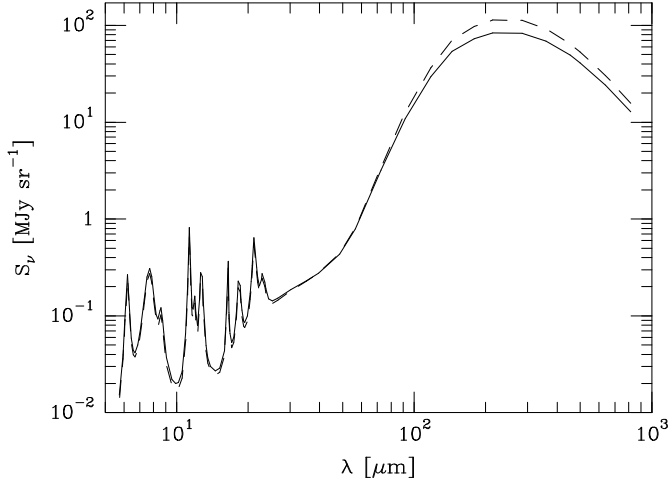}}
\caption[]{Spectra calculated for the coldest core in model cloud $C1$ using
the library method. The libraries are based on simulations with models with
either 32$^3$ (dashed line) or 90$^3$ (solid line) cells. This represents the
overall worst case where the library based on 32$^3$ cell model overestimates
e.g. the IRAS ratio $I_{\rm 60}/I_{\rm 100}$ by more than 10\%.}
\label{fig:dspectra}
\end{figure}

We tested the effect explicitly with model $C1$ discretized into 90$^3$ cells.
The library was built as above (including PAH) but based on simulation of the
radiation field in models with either 32$^3$, 64$^3$ or 90$^3$ cells. Spectra
computed with the different libraries were compared and the results are shown
in Table.~\ref{tab:rms}. The rms difference between the predicted IRAS
intensity ratios, $I_{\rm 60}/I_{\rm 100}$ are below one percent and only in
the coldest core at the position (-2.1, -1.1) the error increases to
$\sim$12\%. The spectra for this position are shown in
Fig.~\ref{fig:dspectra}. This shows that the spectrum of the coldest spot in
the cloud is not necessarily well reproduced when the library is based on a
model with crude discretization.

It is not possible to solve the dust temperature distributions for all 90$^3$
cells in a reasonable time using direct calculations i.e. without the library
method. The dust emission was solved, however, for two layers of cells (i.e.
two times 90$\times$90 cells) that included the coldest core identified from
the computed map of IRAS ratio $I_{\rm 60}/I_{\rm 100}$. Altogether
2$\times$90 spectra were computed for different lines-of-sight along this
slice. Comparison of this subset with spectra computed with the library method
shows rms difference of $\sim$1-2\%, irrespective of the library used. This
indicates that, apart from the coldest spot, the libraries are equally good in
predicting the infrared spectrum. For the library based on simulations with a
90$^3$ cloud the maximum deviation at any single frequency is down to 5\%.

\begin{table}
\caption[]{Comparison of spectra calculated for the model $C1$.
The cloud was discretized into 90$^3$ cells. The libraries were created using
simulations of the radiation field in models with either 32$^3$, 64$^3$ or
90$^3$ cells. The last one was used also as the reference to which all other
models were compared. The rms difference for range 10-300$\mu$m as well as
maximum deviation in IRAS ratio $I_{\rm 60}/I_{\rm 100}$ are
shown. The last column shows the comparison with direct calculations (see
text). }
\begin{tabular}{lllll}
 model    &   library (32$^3$)  & lib. (64$^3$) & lib. (90$^3$) & direct \\
 rms($\Delta I$)                      & 0.4\%  & 0.15\% & -- & 1.3\% \\
 max $\Delta(I_{\rm 60}/I_{\rm 100}$) & 12.4\% & 2.5\%  & -- & 1.4\% \\
\end{tabular}
\label{tab:rms}
\end{table}

The effect of the cloud resolution on the created library is relatively small.
There are at least two ways to reduce these errors further. One can always create
the library using the same discretization as used in the calculation of the
spectra. This can be time consuming, but the library method will still be much
faster than the direct solutions since the dust emission is solved only for
the radiation fields selected in the library and not for all cells. Another
possibility is to create the library using a model where the range of
intensities is artificially increased. This can be achieved by creating the
library using a model cloud with a slightly higher density.

\subsection{Effect of cloud resolution on spectra}

In this section we compare spectra calculated for model $C1$ with cloud
discretized into either $32^3$ or 128$^3$ cells. All calculations are made
using the same library which includes three reference wavelengths, each with
23 intensity intervals. With increasing resolution, the edges of high density
regions are better resolved, and this might lead to changes in the intensity
distributions and/or intensity ratios.

\begin{figure}
\resizebox{\hsize}{!}{\includegraphics{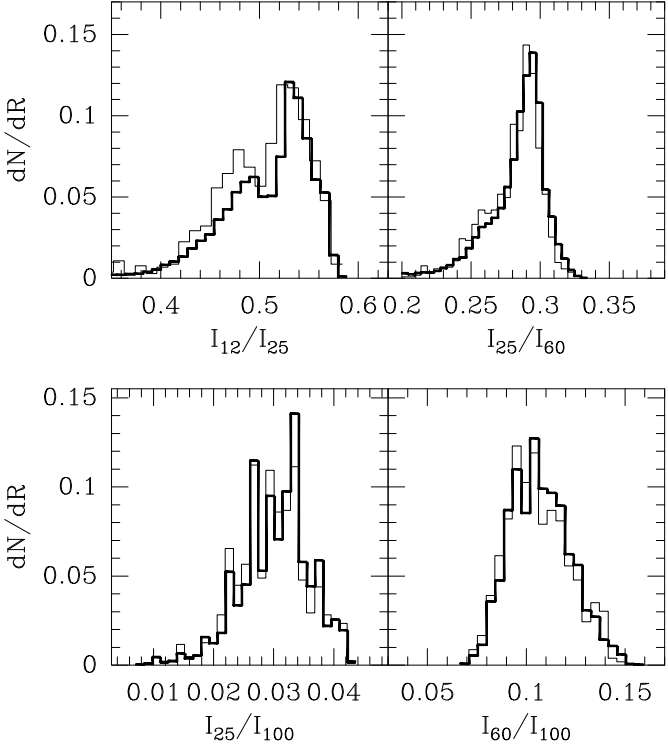}}
\caption[]{Distributions of selected IRAS colour ratios
in maps computed using either 32$^3$ (thin line) or 128$^3$ cell models (thick line).
}
\label{fig:histo_ratio}
\end{figure}

Table~\ref{tab:range} shows the ranges of surface brightness values for
the four IRAS bands for the two discretizations. As expected, with low
resolution, the intensity variations are smoothed and the range of intensities
is also correspondingly reduced. Observationally, this is more important at the
limit of high surface brightness values, where the increase of the spatial
resolution by a factor of four increases the peak intensities by approximately
the same amount. In the low-resolution model, each spectrum corresponds to an
average surface brightness over a larger area, but the values themselves are
not significantly biased. For example, the predicted colour ratios shown in
Fig.~\ref{fig:histo_ratio} are essentially the same irrespective of the
resolution.
The ratio $I_{\rm 12}/I_{\rm 25}$ changes slightly as the resolution is
increased. This ratio traces the PAH component and higher values imply
stronger radiation field. In particular, the values in the higher peak at
$I_{\rm 12}/I_{\rm 25}\sim$0.55 originate mostly from the surface of the
projected cloud.

\begin{table}
\caption{
The range of surface brightness values (Jy sr$^{-1}$) in IRAS bands for models
discretized into either $L^3$=32$^3$ or 128$^3$ cells. The power of ten is
given in parenthesis.}
\begin{tabular}{rllll}
$L$               &  I(12$\mu$m)  & I(25$\mu$m)  & I(60$\mu$m)  & I(100$\mu$m)  \\
\hline
32  &   6.6(4)-1.1(6)   & 1.5(5)-2.5(6) & 4.3(5)-8.5(6)  & 4.8(6)-7.5(7) \\
128 &   5.1(4)-2.0(6)   & 1.1(5)-3.5(6) & 3.4(5)-1.6(7)  & 3.5(6)-1.4(8) \\
\end{tabular}
\label{tab:range}
\end{table}

The average properties of the emission, the surface brightness and colour
ratios are almost independent of the discretization. Compared with the 32$^3$
model, however, the calculation with the higher resolution yields an
approximately 2\% lower total infrared luminosity. Although small, the change
is significant, since the estimated uncertainty per spectrum is already of the
same order. The most likely explanation is that at lower resolution, the holes
that appear between the filaments are smoothed out, and this increases the
total probability that incoming photons are absorbed within the cloud. Bernard
et al. (\cite{bernard92}) noted that for spherically symmetric models with
power law density distributions, the approximation of the radial profile with
a step function leads to {\em under}estimation of the infrared emission. This
effect must also apply to the emission predicted from the cores and filaments
in three-dimensional models, especially as the discretization is much more
coarse by necessity. At least in the present case this did not seem to have
affected the average surface brightness values. However, without a relatively
high resolution, the dense cores would not be resolved at all.

Perhaps the main advantage of high resolution is that it enables statistical
studies of the simulated infrared maps. In such studies the map size should
preferably be at least 256$^2$ in order to provide sufficient range of scales.

\section{MHD models} \label{sect:models}

In this section we present dust emission spectra computed with
three-dimensional model $C1$ consisting of $128^3$ cells. Emission is
calculated for the range 5$\mu$m--300$\mu$m using the dust model of Draine \&
Li (\cite{draine01}). The density field is the solution of the
three-dimensional MHD equations, in a regime of forced highly supersonic and
super-Alfv\'{e}nic turbulence and with an isothermal equation of state (model
$C$ in Juvela et al. ~\cite{juvela01}). We first present results from the
original model and compare them with calculations with lower discretization
models. We also discuss the effect of the cloud structure on the observed
surface brightness maps and, finally, study qualitatively the effects
introduced by spatially varying dust abundances.

\subsection{The original model}

The library (i.e. the mapping between the incoming intensity at selected
reference wavelengths and the resulting emission) is created using the
original models resampled onto a $32^3$ cell grid. We use three reference
wavelengths, each with 23 intensity intervals. As described above, the
downsized model is needed only to estimate the range of intensities and
profiles for the incoming radiation. Therefore, at this stage the spatial
resolution is not crucial for the final accuracy. The use of the library leads
to significant time saving since in the large, 128$^3$ cell model the
radiative transfer problem is solved only at the three reference wavelengths.
In this way the computational time is reduced by a factor of $\sim$100 (ratio
of total number of simulated wavelengths and number of reference wavelengths).
The main cost lies, however, in the solving of the dust temperature
distributions. Here the time saving for the $128^3$ cloud is a factor of
$128^3/23^3\sim$170.  
As shown in Sect.~\ref{sect:tests}, the expected rms
error in the computed spectra is less than 2\%.

\begin{figure}
\resizebox{\hsize}{!}{\includegraphics{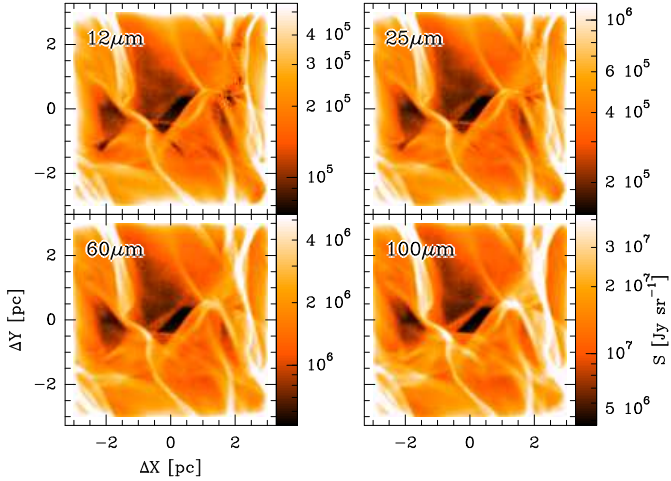}}
\caption[]{
Surface brightness in the four IRAS bands computed for the model $C1$.}
\label{fig:orig_maps}
\end{figure}

\begin{figure}
\resizebox{\hsize}{!}{\includegraphics{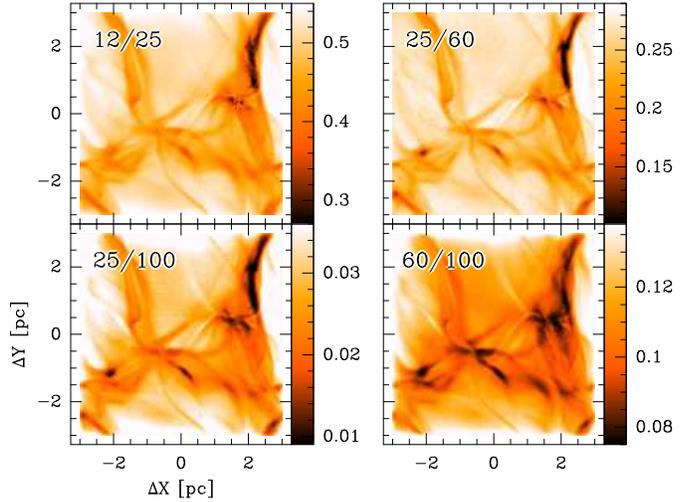}}
\caption[]{
Ratios between surface brightness values in IRAS bands.
}
\label{fig:ratio_maps}
\end{figure}

The surface brightness maps for the four IRAS bands are shown in
Fig.~\ref{fig:orig_maps}. The maps have been calculated using the bandpass
profiles given in the IRAS Explanatory Supplement (Beichmann et al.
\cite{beichmann88}). In all four maps the surface brightness probes 
the same filamentary density field. Maps of surface brightness ratios
(Fig.~\ref{fig:ratio_maps}) show, however, clear variations in the FIR colours
and, as expected, dense cores and filaments are clearly visible through their
lower colour temperatures.

\begin{figure}
\resizebox{\hsize}{!}{\includegraphics{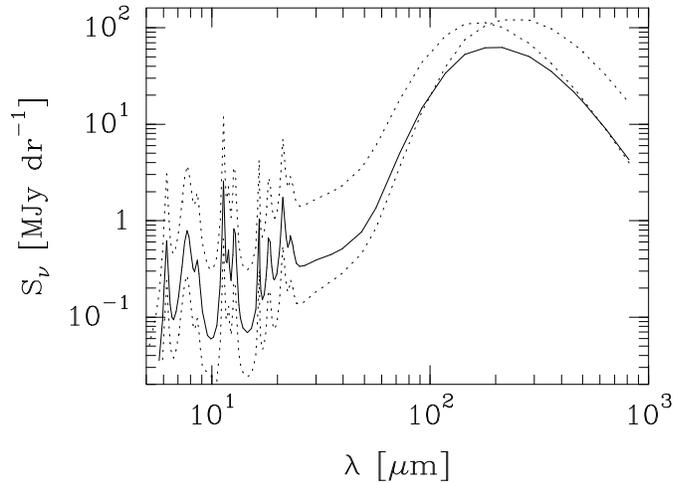}}
\caption[]{
Average infrared spectrum observed from model $C1$. The dotted lines show two
individual spectra with extreme ratios between 15$\mu$m and 600$\mu$m
intensities. }
\label{fig:spectra}
\end{figure}

Figure~\ref{fig:spectra} shows the average infrared spectrum and two
individual spectra with extreme values of the ratio between 15$\mu$m and
600$\mu$m intensities. The spectrum with relatively strongest mid-infrared
emission originates from the upper left hand corner i.e. gas subjected almost
to the full intensity of the external radiation field. The coldest spectrum is
observed at the core located at (-2.1, -1.1). There the emission peaks close
to $\ga$200$\mu$m but the core is already quite noticeable in the maps of IRAS
colour ratios (Fig.~\ref{fig:ratio_maps}).

For the Li \& Draine (\cite{li01}) dust model, the ratio $I_{\rm 60}/I_{\rm
100}$ is close to 0.2 when dust is exposed to the pure interstellar radiation
field.
The dust model of D\'esert et al. (\cite{desert90}) predicts the same ratio
$I_{\rm 60}/I_{\rm100}\sim$0.2. This is, of course, not surprising since both
dust models are created to the reproduce existing IRAS and COBE measurements.
The ratio of 0.2 is similar to those found towards many diffuse clouds (e.g.
Laureijs et al. \cite{laureijs91}). This ratio applies, however, only to the
diffuse medium. In our cloud $C1$, the average ratio is $\sim$0.11. This cloud
is quite opaque, with mean visual extinction of $A_{\rm V}\sim$11. The model
$C0$ has lower visual extinction, $\langle A_{\rm V}\rangle\sim$2.8, and is
closer to typical regions of molecular clouds. In $C0$, the average colour
ratio is $\langle I_{\rm 60}/I_{\rm 100}\rangle
\sim$0.15. This is the same as the value found by Abergel et el.
(\cite{abergel94}) for the Taurus molecular cloud complex.

Some of the lowest colour temperatures are reached in the filament in the
upper right hand corner of the maps. Like many other features, this is
actually a superposition of separate filaments or sheet-like structures along
the line of sight. In this direction, the densest filament is $\sim$1\,pc in
diameter and has a maximum density of $\ga$10$^4$\,cm$^{-3}$. This roughly
corresponds to a visual extinction of 10$^{\rm m}$ across the filament
($\sim$5$^{\rm m}$ to the centre of the filament). There is, however,
significant additional extinction from the more diffuse material between the
dense regions. The ratio $I_{\rm 60}$/I$_{\rm 100}$ decreases within the filament by a
factor of $\sim$2.0 relative to the average value over the map, and is
therefore approximately 4 times lower than the value given by the dust model
in the diffuse medium. In observations the ratio is often found to vary by a
larger amount. In Lynds~134 the ratio decreases by a factor of $\sim$7 from
the diffuse medium to the cloud cores, where the visual extinction is
$A_{\rm V}\sim$10, close to that in our model $C1$ (Laureijs et
al.\cite{laureijs91}). In the cirrus cloud MCLD\,123.5+24.9 in the Polaris
Flare the observed ratio $I_{\rm 60}/I_{\rm 100}$ is 0.034 (Bernard et al.
\cite{bernard99}) while the quoted average value for cirrus clouds is $\langle
I_{\rm 60}$/I$_{\rm 100}\rangle\sim$0.31) (Lagache et al. \cite{lagache98}). The ratio
$\langle I_{\rm 60}$/I$_{\rm 100}\rangle$ has therefore decreased by a factor of
$\sim$9, although the extinction is lower than in the filament in our model.
As discussed by Bernard et al. (\cite{bernard99}), the explanation of the
observations requires spatial variation of the dust properties. We will study
these effects qualitatively in Sect.~\ref{sect:adjust}.
The calculations were carried out using a cloud divided into 128$^3$ cells
while the library was created with a model with only 32$^3$ cells.  However,
according to Sect.~\ref{sect:tests} the expected rms error of the spectra is
only $\sim$2\% and the error in $I_{\rm 60}/I_{\rm 100}$ is less than
15\%.

The average ratio $\langle I_{\rm 140}/I_{\rm 240} \rangle$ calculated using
the DIRBE system response functions is $\sim$0.88 for model $C1$ and
$\sim$1.05 for model $C0$. In the spherically symmetric models of Bernard et
al. (\cite{bernard92}; see Lagache et al.\cite{lagache98}, Table~3), similar
ratios are obtained only when the extinction is low compared with the average
extinction in our models. For example, with $A_{\rm V}$=4 towards the cloud
centre, their ratio has already decreased to 0.77, well below the value in
model $C1$ which has, nevertheless, higher average visual extinction. This may
be partly caused by differences in the dust models, but it is mainly the
result of the inhomogeneity of our models. In an inhomogeneous cloud, the ratio
remains higher because there are no large, connected regions completely
shielded from the external radiation field.

\begin{figure}
\resizebox{\hsize}{!}{\includegraphics{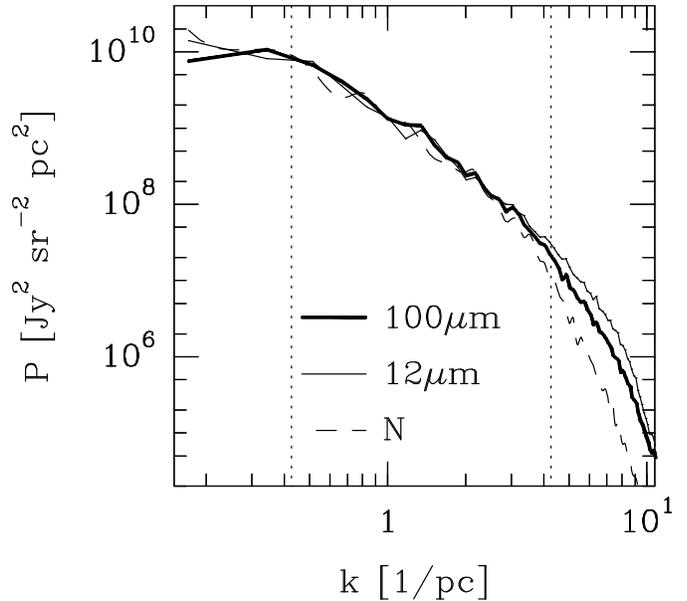}}
\caption[]{
Spatial power spectra of the 12$\mu$m (thin line) and 100$\mu$m (thick line)
surface brightness maps computed from model $C1$. The 100\,$\mu$m curve has
been scaled down by a factor of 5000. The Nyquist limit corresponds to
$k=10.7$\,pc$^{-1}$, but since the actual resolution of the MHD simulation is
somewhat lower, the power already drops above $k\sim$5\,pc$^{-1}$. The dashed
curve shows the power spectrum of the underlying column density map with
arbitrary scaling. The vertical lines limit the range used to compute the
exponents of the corresponding power laws.}
\label{fig:power_spectra}
\end{figure}

The spatial power spectra for the 12$\mu$m and 100$\mu$m maps are shown in
Fig.~\ref{fig:power_spectra}.
These are one-dimensional power spectra, i.e. they are computed from the
corresponding two-dimensional power spectra by averaging over annuli of
constant wavenumber.
Although the MHD simulation used periodic boundary conditions, the external
heating introduces radial gradients in the infrared maps, and therefore direct
comparison with observations is not possible at the largest scales. We will
address this question in future work, where more detailed modelling will be
done, also taking into account the asymmetry of the galactic radiation field.

The finite spatial resolution of the MHD simulation and the relatively small
size of the maps, 128$\times$128 pixels, further limits the available dynamic
range. In Fig.\ref{fig:power_spectra} the vertical lines show the range where
the spectrum follows an approximate power law, and where we determine the
exponent of the power law. We obtain exponents $\beta=$-2.54 and -2.72 for
12$\mu$m and 100$\mu$m, respectively.
In Sect.~\ref{sect:tests} we concluded the accuracy of the computed spectra is
probably of the order of 2\% per frequency point. Averaged over the IRAS
filters the noise should be much lower. This is confirmed by the fact that the
power spectra do not show a flat part typical of noise. Assigning an
uncertainty of 2\% to the computed IRAS values we estimated with Monte Carlo
methods an uncertainty of $\sim$0.01 for the exponent of the power law. This
error estimate is valid for the comparison of maps with the same underlying
density distribution. The exponent could be changed significantly more by
selecting a different range of spatial frequencies or e.g. a different
projection of the cloud. 

The power law exponent for the column density map is -2.80.
The same power spectrum slope is predicted analytically for the projected
density in supersonic turbulence by Boldyrev et al. (\cite{boldyrevetal02}), 
based on the velocity structure function scaling 
predicted by Boldyrev (\cite{boldyrev02}).

Compared with the column density maps the infrared maps have more power on
small scales but the slopes are in the range obtained from observations. The
power law exponent of IRAS maps determined by Gautier et al.
(\cite{gautier92}) was $k=$-3.0. Based on ISO observations, however,
Herbstmeier et al. (\cite{herbstmeier}) found different power laws
in different clouds, from very shallow spectra, $k\sim -1$, in dark regions up
to $k=$-3.6 in the Cepheus flare.

The power law exponent of the infrared maps is mostly a result of the
underlying density distribution. The infrared maps do contain, however,
significantly more power at high spatial frequencies, above the range used to
compute the values of the power law exponents. This is due to dust temperature
variations that cause intensity fluctuations on top of the column density
distribution. The 12$\mu$m map seems to have more small scale structure than
the 100$\mu$m one.
At these high frequencies the underlying density distribution is affected
by the finite resolution of the MHD simulations. However, for the given
density distribution we can compare the power spectra up to the Nyquist limit
i.e. in Fig.~\ref{fig:power_spectra} up to $\sim$10\,kpc$^{-1}$. The
differences were also observed in other models and they are well above the
Poisson noise from the Monte Carlo simulation. Most of the fluctuations in the
power spectra are caused by the underlying, common density distribution and
e.g. the difference 12\,$\mu$m and 100\,$\mu$m is more significant than a
comparison with the apparent noise in Fig.~\ref{fig:power_spectra} would lead
to believe.
The quantitative result could, however, change with better spatial resolution
if the density field contained significant small scales structures.

At 12$\mu$m, the emission mostly depends on the heating by short
wavelength photons that are absorbed at the surface of the density structures.
On the other hand, the heating of the large grains varies less rapidly, and one
could expect the 100$\mu$m map to show relatively less small scale structures.
Nevertheless, analysis of IRAS maps has not shown clear differences between
the different bands. It is conceivable that in the observations, the
differences are masked by projection effects or by the spatial variation in
dust abundances and properties.

\subsection{Effects of inhomogeneity}

To check to what extent the three-dimensional cloud structure affects the
properties of infrared emission, we compute spectra from two modified models,
one with a smoother density structure and one with a smaller density
contrast than model $C1$. 

\begin{figure}
\resizebox{\hsize}{!}{\includegraphics{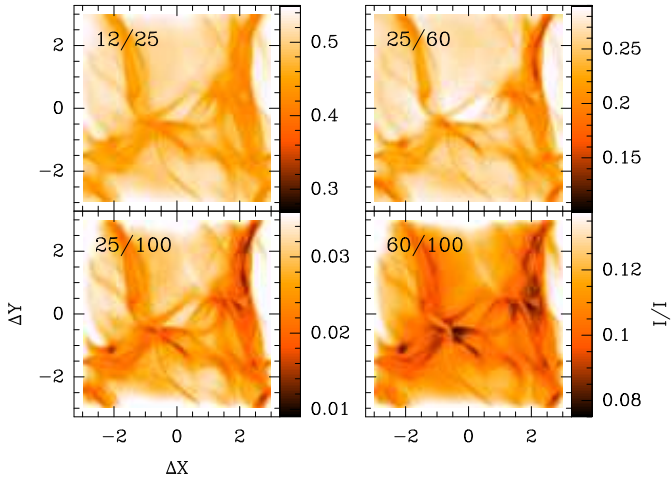}}
\caption[]{IRAS colours for model $C1a$, a modification of model $C1$
where the density structure along the line of sight has been made smooth by
re-arranging the cells along the line of sight (see text). 
}
\label{fig:smooth_maps}
\end{figure}

In the case of the first model, $C1a$, we re-arrange the cells in model $C1$
so that in each column along the line of sight, the density increases
monotonically towards a plane that is perpendicular to the line-of-sight and
goes through the cloud centre. In this way, both the column density field and
the probability density function of the volume density are the same as in the
original model C1, but the density structure is smoother. Comparison with the
original model showed some noticeable changes in the surface brightness
distribution, which now follows more closely the column density distribution.
In the original cloud, the observed infrared intensities are more sensitive to
the intensity of the radiation field at the location of each individual
filament or core. For example, the filament going through position (-2,2) is
brighter in Fig.~\ref{fig:orig_maps} than in Fig.~\ref{fig:smooth_maps}
because it is located near the surface of the cloud.

In model $C1a$, there are also some changes in the distribution of IRAS
colours (see Fig.~\ref{fig:smooth_maps}). There is a tighter correlation
between 60$\mu$m and 100$\mu$m intensities than in the original model $C1$.
This is expected, as the cloud is inhomogeneous (clumpy or filamentary)
essentially only in the plane of the sky, and smooth in the third spatial
dimension.

The modified cloud $C1$ is not `clumpy' along the line of sight, and following
the preceding discussion on the effects of resolution, one might expect higher
infrared emission. However, the emission is some 20\% {\em lower}.  This
is a direct consequence of the fact that the modified cloud is effectively
smaller along the line of sight. Since most of the mass is concentrated to a
smaller volume, less radiation enters dense parts of the cloud and the total
absorbed energy is also correspondingly lower. Conversely, if dust properties
and the external radiation field were known to an accuracy of a few percent,
it would be possible to make rough estimates of the line-of-sight size of
clouds, even when the degree of clumpiness in this direction is not known.

In model $C1$, the ratio between the highest and lowest densities is close to
6$\times 10^4$. We produced another, more uniform cloud model, $C1b$.
By applying a linear scaling to the density values, the total density contrast
was reduced by a factor of 100. 
The average column density remains the same as in models $C1$ and $C1a$, but
the absorbed energy is higher since it depends on column density via
an exponential function. As a result, compared with the original model $C1$, the
total emitted energy increases by $\sim$7\%. For the IRAS bands, the difference
is slightly higher, $\sim$8.5\%. In spite of the increased energy input, the
average ratio between IRAS bands gives a {\em lower} colour temperature. The
difference is probably real, although it is only a couple of per cent, which is
close to the expected accuracy of the calculations. In model $C1b$, near- and
mid-infrared emission is concentrated on the surface of the model cloud, and
the inner parts are correspondingly cooler. The infrared colours were averaged
over all spectra without weighting them with the corresponding intensities.
Consequently, the average colours can indicate lower dust temperatures than in
the model $C1$, where the short wavelength emission is spread more evenly over
the projected cloud area.

The power spectra of the 12$\mu$m and 100$\mu$m maps are almost identical in
the range (see Fig.~\ref{fig:power_spectra}) used for the determination of the
power law exponent and the exponent is $\sim$-2.7 for both maps. If the power
spectra are scaled on top of each other in this range of spatial frequencies,
the 12$\mu$m curve is only slightly above the 100$\mu$m curve at high
frequencies, while at low frequencies, the 12$\mu$m is significantly higher,
due to the stronger brightening towards the cloud surface.

\subsection{Dust abundance variations} \label{sect:adjust}

There is strong observational evidence that the dust properties change, not
only in dense cloud cores (e.g. Laureijs et al. \cite{laureijs96}) but also in clouds with
moderate visual extinction (Bernard et al. \cite{bernard99}; Cambr\'esy et al.
\cite{cambresy01}; Stepnik et al. \cite{stepnik_esa}; Miville-Desch\^enes et
al. \cite{miville02}). Possible causes are the growth of grain mantles and the
formation of large dust aggregates. The processes can take place only in dense
medium that is protected from the external UV radiation field.

We use the three-dimensional model $C1$ to study the {\em qualitative}
effects of abundance changes on the observed infrared radiation. The
model has mean density $\sim 10^3$\,cm$^{-3}$ with peak values higher by two
orders of magnitude.

We adjust the abundances of the grain populations according to the local
density only. The modified model is called $C1c$. At each position, the dust
is composed of five components, the first three of which are PAH, silicate and
carbon grains of the Li \& Draine model (\cite{li01}). The additional two
components are modified silicate and carbon grain populations with a flatter
size distribution: the original exponent of the size distribution is adjusted
so that the original number ratio of grains below and above 25\,nm is reduced
by 50\%. At low densities, the abundance of the new components is zero. Their
abundance increases linearly with the logarithm of the density. The
abundance of the unmodified grain populations is decreased linearly with
respect to the logarithm of density and the total mass of dust remains
unchanged. Above 10$^4$\,cm$^{-3}$, all dust is in the modified silicate and
carbon grain populations. This is, of course, only a qualitative description
of the expected dust abundance variations. In particular, we ignore all
changes in the dust optical properties expected to be caused by the formation
of aggregate grains. As a result, the model can be expected to underestimate
true colour temperature variations across dense filaments or cores (see e.g.
Stepnik et al. \cite{stepnik_esa}).

\begin{figure}
\resizebox{\hsize}{!}{\includegraphics{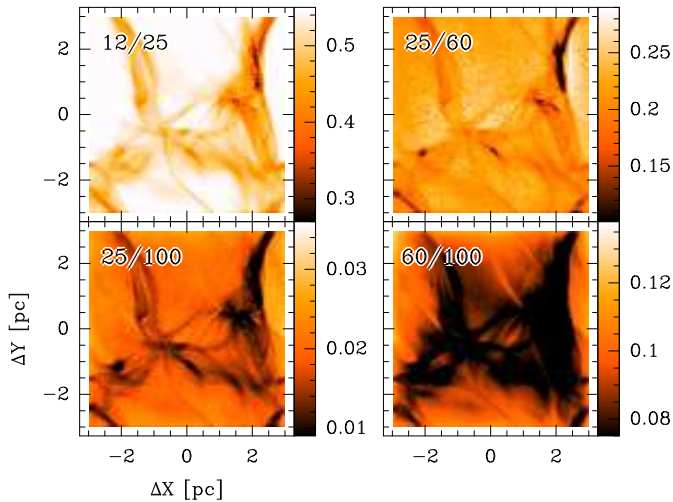}}
\caption[]{
Maps of selected IRAS colours calculated for the model $C1c$. In the
model the dust abundances vary depending on the local density (see text). 
The colour scales are the same as in Fig.~\ref{fig:ratio_maps}.
}
\label{fig:abundance_maps}
\end{figure}

Figure~\ref{fig:abundance_maps} shows maps of IRAS intensity ratios in model
$C1c$. Since we have reduced the number of small grains the average colour
temperatures have dropped. The ratio $\langle I_{\rm 25}/I_{\rm 100} \rangle$
has decreased the most, by 34\%. The ratio $\langle I_{\rm 60}/I_{\rm 100}
\rangle$ is now lower by 23\%. The average 100$\mu$m surface brightness has
only increased by $\sim$17\% and the 60$\mu$m emission is below that
of model $C1$.

The 12$\mu$m map contains relatively more power at small scales. The short
wavelength maps are not expected to change much, since most of the emission
comes from low density material where grain abundances remain almost
unchanged. Abundance variations should increase far-infrared emission in the
densest regions, and the corresponding maps are, therefore, expected to
contain more power at high spatial frequencies. However, the changes in the
power spectra remain very small. The new exponents of the power laws are -2.51
and -2.67 for 12$\mu$m and 100$\mu$m, respectively; the 100$\mu$m slope has
only flattened by $\sim$2\% relative to the original model $C1$.

In the filament at position (2,2) the 12$\mu$m surface brightness has
decreased by $\sim$30\%, while the 100$\mu$m emission has increased less than
10\%: this is {\em less} than the average increase over the whole map. The
change in the dust abundances leads to lower extinction, especially at short
wavelengths, and this decreases temperature variations across the filament.
The lowest values of the ratio $I_{\rm 60}/ I_{\rm 100}$ are $\sim$0.05 in
this region, i.e. some 30\% lower than in the original cloud $C1$. Compared
with the diffuse medium value of 0.2, the reduction in $I_{\rm 60}/ I_{\rm
100}$ is still only a factor $\sim$4. This is still less than found in many
observations (e.g. Laureijs et al. \cite{laureijs91}; Bernard et al.
\cite{bernard99}). The model cloud has very significant extinction, even
outside the densest filaments, and this should help to further decrease the
$I_{\rm 60}/ I_{\rm 100}$ ratio.

In the presented model, we have only adjusted the grain size distributions by
halving the ratio of grains below and above 25\,nm in dense cloud regions with
$n>10^4$\,cm$^{-3}$. There is some evidence that dust properties can undergo
significant changes even at much lower densities, below $10^3$\,cm$^{-3}$
(e.g. Bernard et al. \cite{bernard99}). On the other hand, sub-mm observations
of dense cores seem to indicate that in those environments, {\em all} small
grains may be removed (e.g. Stepnik et al. \cite{stepnik_esa}). If small
dust grains stick onto the surfaces of larger grains, the optical properties
of the grains will change and lead to a lower physical temperature. As a
result, the far-infrared colour temperatures decrease, and this can explain
the very low $I_{\rm 60}/ I_{\rm 100}$-ratios observed. In dense environments,
dust may also form larger, fluffy aggregate particles that have similarly
enhanced far-infrared emissivity and low physical temperature (Mathis \&
Whiffen \cite{mathis89}; Ossenkopf
\cite{ossenkopf93}; Wright \cite{wright87}; Bazell \& Dwek \cite{bazell90}).
Many of the observed cold cores are too small for the low colour
temperatures to be explained merely by the attenuation of the external
radiation field. The need for large variations in the abundance ratio of large
and small dust grains is, therefore, already well established.

\section{Conclusions}

We have presented a new approximate method that can be used to compute the
infrared dust emission from large, three dimensional cloud models consisting
of millions of computational cells. The method is based on simple discretization of the
incoming intensity using a few reference wavelengths. We have demonstrated
that relative accuracy of a few percent is easily reached, even when
transiently heated grains are included. Such accuracy is hard to match
either by observations or the Monte Carlo sampling that is often used to
estimate the intensity of the radiation field. We have tested the method using
model clouds with average visual extinctions up to $A_{\rm V}\sim$100. However,
with use of more reference wavelengths and/or higher discretization, the method
can be applied to even denser clouds, and to models with several heating sources
with different emission spectra.

We have used our new approximate method to compute dust emission, from
near-infrared to sub-mm wavelengths, from models based on MHD simulations with
supersonic and super-Alfv\'enic turbulence. The spatial discretization was
found to have little impact on the derived distributions of the intensity
ratios between IRAS bands. On the other hand, spatial averaging will always
reduce the range of intensities. High resolution calculations are therefore
necessary, especially when the computed infrared maps are used for statistical
studies or for direct comparison with observations.

The spatial power spectra at different wavelengths generally show little
variation. The shorter wavelengths have, however, slightly higher relative
power at higher spatial frequencies. For the selected MHD model, the power
spectra follow approximately a power law with an exponent between -2.5 and -2.7,
while the power spectrum of the underlying column density map has a slope
of -2.8.

We qualitatively studied the effect of density dependent dust size
distributions on the observable infrared maps. By reducing the relative
abundance of very small grains by 50\% in regions with densities
$n>10^4$\,cm$^{-3}$, the ratio $I_{\rm 60}/ I_{\rm 100}$ could be brought down
to $\sim$0.05 in dense regions where visual extinction is $A_{\rm V}\sim$10.
In observations, the colour ratio can reach lower values even in less opaque
clouds. This indicates that dust properties must undergo even larger changes,
with the removal of most of the small grains and the growth of larger ones.

The numerical method presented in this work will be very useful to constrain
dust properties by comparison with new observational data. Dust column density
maps of large molecular cloud complexes can now be obtained from stellar
extinction measurements based on the ``Two Microns All Sky Survey'' (2MASS).
Padoan et al. (\cite{padoan2002}) have recently generated
extinction maps of the whole Taurus molecular clouds complex using 2MASS data,
with a spatial resolution comparable or better than that of 100~$\mu$m IRAS
images, spanning a range of visual extinction from $A_{\rm V}\approx 0.3$~mag
to $A_{\rm V}\approx 30$~mag. The comparison of this type of dust column
density maps with IRAS images and, in the near future, with SIRTF-MIPS images
will provide new observational constraints over a large range of column
density and gas density values. Detailed three dimensional calculations of
dust emission from physical models of interstellar clouds will be necessary
for reproducing the observational data and improve our knowledge of the
properties of interstellar dust.

\begin{acknowledgements}
We thank the referee, Dr. J. Le Bourlot, for comments that have improved the
paper.
M.J. acknowledges the support of the Academy of Finland Grants no. 1011055,
166322, 175068, 174854. The work of P.P. was performed while P.P. held a
National Research Council Associateship Award at the Jet Propulsion
Laboratory, California Institute of Technology.
\end{acknowledgements}

\end{document}